\documentclass[prd,aps,eqsecnum,showpacs,nofootinbib]{revtex4}

\usepackage{amsmath,amssymb}
\usepackage{bm}
\usepackage[dvips]{graphicx,color}

\begin{document}

\title{Conformal equivalence in classical gravity:\\
the example of ``veiled" General Relativity}

\author{Nathalie Deruelle$^1$ and Misao Sasaki$^{2,3}$}

\affiliation{
$^1$APC, UMR 7164 du CNRS, Universit\'e Paris 7, 
75205 Paris Cedex13, France
\\
$^2$Yukawa Institute for Theoretical Physics, Kyoto University, 
Kyoto 606-8502, 
Japan
\\
$^3$Korea Institute for Advanced Study
207-43 Cheongnyangni 2-dong, Dongdaemun-gu, 
Seoul 130-722, Republic of Korea
}

\date{July 21, 2010}

\begin{abstract}
In the theory of General Relativity, gravity is described by a metric
 which couples minimally to the fields representing matter.
 We consider here  its ``veiled" versions  where the metric is conformally
 related to the original one and hence is no longer minimally coupled to 
the matter variables. 
We show on simple examples that 
observational predictions are nonetheless exactly the same as in General 
Relativity, with the interpretation of this
``Weyl" rescaling ``\`a la Dicke", that is,
as a spacetime dependence of the inertial mass of the matter
constituents.
\end{abstract}

\hfill YITP-10-61

\maketitle

\section{INTRODUCTION}

Many extensions of General Relativity which are under current investigation 
(for example $f(R)$ gravity, see e.g. \cite{Tsujikawa10}, or quintessence 
models, see e.g. \cite{Copeland06}) fall in the class of scalar-tensor 
theories (see e.g. \cite{Maeda02}) where gravity is represented by a scalar 
field $\tilde\phi$ together with a metric $\tilde {\bm g}$ which minimally 
couples to the matter variables. Now, as is well-known (see \cite{Damour92} 
 where references to the earlier literature can also be found), the ``Jordan 
 frame" variables $\tilde\phi$ and $\tilde{\bm g}$ can be traded for the
 ``Einstein frame" variables ($\phi_*, {\bm g}_*$) with 
$\tilde{\bm g}=e^{2\Omega}{\bm g}_*$, the conformal factor $\Omega$
 being chosen so that the action for gravity becomes Einstein-Hilbert's, 
the ``price to pay" being that the matter fields no longer minimally 
couple to the metric ${\bm g}_*$.

Although there seems to be an agreement in the recent literature about 
the mathematical equivalence of these two ``frames" (as long as $\Omega$ 
does not blow up) there is still some debate about their ``physical" 
equivalence, the present trend (see e.g. \cite{Tsujikawa10} and references
 therein)  being that calculations may be performed in the Einstein frame
 but interpretation should be done in the Jordan frame (for the opposite
 view see e.g. \cite{Gunzig98}, where a comprehensive review of the earlier 
literature can also be found).

It should be clear however, see \cite{Dicke62}, that, just as one can 
formulate and interpret a theory using any coordinate system (proper
 account being taken of inertial accelerations if need be), one should
 be able to formulate and interpret (classical) gravity using any
 conformally related metric, proper account being taken of non-minimal 
coupling if need be. (For recent papers supporting this view, 
see e.g. \cite{Flanagan04,Catena06,Faraoni06}.)

In this paper we shall try to make this equivalence ``crystal 
clear" by showing that some familiar predictions of General Relativity 
can equivalently be made in its ``veiled" versions where the metric 
is conformally related to the original one and hence is no longer 
minimally coupled to the matter variables.

\section{Conformal transformations and ``veiled" General Relativity}

In the theory of General Relativity:
\begin{list}{--}{}
\item Events are represented by the points $P$ of a 4-dimensional  
manifold ${\cal M}$ equipped with a Riemannian metric ${\bm g}$, 
with components $g_{\mu\nu}(x^\alpha)$
 in the (arbitrary) coordinate system $x^\alpha$ 
labelling the points $P$. 

\item
 Matter is represented by a collection of tensorial fields on ${\cal M}$,
 denoted $\psi_{(a)}(P)$. 

\item
 Gravity is encoded in the metric ${\bm g}$ which couples minimally to
 the fields $\psi_{(a)}$. This means that the action for matter is obtained 
from the form it takes in flat spacetime in  Minkowskian coordinates by 
replacing  $\eta_{\mu\nu}$ by $g_{\mu\nu}$.

\item
Finally the action for gravity is postulated to be Einstein-Hilbert's. 
\end{list}
Hence the familiar total action:
\begin{equation}
S[g_{\mu\nu},\psi_{(a)}]
={1\over16\pi}\int d^4x\sqrt{-g}R+S_m[g_{\mu\nu}, \psi_{(a)}]\,,
\label{HilbertAction}
\end{equation}
  where $g$ is the determinant of the metric components $g_{\mu\nu}$ and $R$ 
the scalar curvature. Our conventions are: signature $(-+++)$, 
$R=g^{\mu\nu}R_{\mu\nu}$, $R_{\mu\nu}=R^\sigma{}_{\mu\sigma\nu}$, 
$R^\mu{}_{\nu\rho\sigma}=\partial_\rho\Gamma^\mu{}_{\nu\sigma}+\cdots$. 
We use Planck units where $c=\hbar=G=1$.
    
The field equations are obtained by extremising $S$ with respect to 
the metric $g_{\mu\nu}$ and the matter fields $\psi_{(a)}$, which yields 
the equally familiar Einstein equations,
\begin{equation}
G_{\mu\nu}=8\pi T_{\mu\nu}\,,\qquad {\delta S_m\over\delta\psi_{(a)}}=0\,,
\label{EinsteinEq}
\end{equation}
where $G_{\mu\nu}=R_{\mu\nu}-{1\over2}g_{\mu\nu}R$ is the Einstein tensor 
and where $T_{\mu\nu}=-{2\over\sqrt{-g}}{\delta S_m\over\delta g^{\mu\nu}}$ 
is the total stress-energy tensor. As is well-known 
 $T_{\mu\nu}$ is constrained by the Bianchi identity to be divergence-less,
\begin{equation}
D_{\nu}T^{\mu\nu}=0\,,
\label{Bianchi}
\end{equation}
 $D$ being the covariant derivative associated with ${\bm g}$.
 Recall that this conservation law implies that the worldline of
 uncharged test particles are represented by geodesics of 
the metric ${\bm g}$.
 \\

Let us now equip our manifold ${\cal M}$ with another metric $\bar{\bm g}$,
 with components $\bar g_{\mu\nu}$ in the same coordinate system $x^\alpha$,
 which is conformally related to the original one:
\begin{equation}
g_{\mu\nu}=\Phi\, \bar g_{\mu\nu}\,,
\label{ConfTransfo}
\end{equation}
$\Phi(x^\alpha)$ being an arbitrary function of the coordinates,
 that we shall restrict to be  everywhere 
positive.\footnote{$({\cal M},\rm{g})$ or $({\cal M},\rm{\bar g})$ 
are often called, rather improperly,  ``frames", when a more accurate 
wording would be ``representations" of space and time,
 see \cite{Dicke62}.\\}  

Using the fact that $\sqrt{-g}=\Phi^2\sqrt{-\bar g}$ and that
 the Ricci tensors and scalar curvatures are related as
\begin{equation}
R_{\mu\nu}=\bar R_{\mu\nu}-{\bar D_{\mu\nu}\Phi\over\Phi}
-{\bar g_{\mu\nu}\over2}{\bar\square\Phi\over\Phi}
+{3\over2}{\partial_{\mu}\Phi\partial_{\nu}\Phi\over\Phi^2}
\,,
\quad 
R={1\over\Phi}\left(\bar R-3{\bar\square\Phi\over\Phi}
+{3\over2}{(\bar\partial\Phi)^2\over\Phi^2}\right)\,,
\end{equation}
($\bar R_{\mu\nu}$, $\bar R$ and $\bar D$ being the Ricci tensor, 
the scalar curvature and the covariant derivative associated with 
the metric $\bar{\bm g}$), it is easy to find the ``veiled" version
 of Einstein's equations (\ref{EinsteinEq}),
\begin{equation}
\Phi\,\bar G_{\mu\nu}-\bar D_{\mu\nu}\Phi+\bar g_{\mu\nu}\bar\square\Phi
+{3\over2\Phi}\left(\partial_{\mu}\Phi\partial_{\nu}\Phi
-\frac{1}{2}\bar g_{\mu\nu}(\bar\partial\Phi)^2\right)
=8\pi\bar T_{\mu\nu}\,,
\quad {\delta S_m\over\delta\psi_{(a)}}=0\,,
\label{veiledEinstein}
\end{equation}
where $S_m$ is now expressed in terms of 
$\bar g_{\mu\nu}$, $S_m[g_{\mu\nu},\psi_{(a)}]=S_m[\Phi\,\bar g_{\mu\nu},\psi_{(a)}]$,
 and where 
$\bar T_{\mu\nu}
=-{2\over\sqrt{-\bar g}}{\delta S_m\over\delta \bar g^{\mu\nu}}$ 
so that $\bar T_{\mu\nu}=\Phi\, T_{\mu\nu}$, with $g_{\alpha\beta}$
 replaced by $\Phi\,\bar g_{\alpha\beta}$ in $T_{\mu\nu}$.
 As for the Bianchi identity (\ref{Bianchi}), it translates into
\begin{equation}
\bar D_{\nu}\bar T^{\mu\nu}={\bar\partial^{\mu}\Phi\over2\Phi}\,\bar T\,.
\label{veiledBianchi}
\end{equation}
The total stress-energy tensor is no longer conserved.

Equations (\ref{veiledEinstein}), (\ref{veiledBianchi}) can also
 be straightforwardly obtained from the Einstein-Hilbert
 action (\ref{HilbertAction}). Indeed it reads, 
using (\ref{ConfTransfo}) and up to a boundary term,
\begin{equation}
S[\bar g_{\mu\nu},\Phi, \psi_{(a)}]
={1\over16\pi}\int d^4x\sqrt{-\bar g}
\left(\Phi\,\bar R+{3\over2}{(\bar\partial\Phi)^2\over\Phi}\right)
+S_m[\Phi\,\bar g_{\mu\nu}, \psi_{(a)}]\,.\label{veiledAction}
\end{equation}
Extremisation with respect to $\bar g_{\mu\nu}$ and $\psi_{(a)}$
 yields (\ref{veiledEinstein}). As for the extremisation
 with respect to $\Phi$ it is redundant since it turns out
 to be equivalent to the trace of  equation (\ref{veiledEinstein}).
 This reflects the fact that,  $\bar g_{\mu\nu}$
 remaining unconstrained,  $\Phi$ is an arbitrary function
 and not a dynamical field.\footnote{One notes the resemblance
 of the action (\ref{veiledAction}) and the field 
equations (\ref{veiledEinstein}) with the Brans-Dicke
 action and field equations \cite{BransDicke61} when
 their parameter $\omega$ is $\omega=-3/2$, see e.g. \cite{Dabrowski05}.
 The difference (which makes $\omega=-3/2$ Brans-Dicke theory 
different from General Relativity) is that, in Brans-Dicke theory,
 matter is minimally coupled to the metric ${\rm\bar g}$ (not ${\bm g}$),
$$
S_{\rm BD}^{\omega=-3/2}
={1\over16\pi}\int d^4x\sqrt{-\bar g}
\left(\Phi\,\bar R+{3\over2}{(\bar\partial\Phi)^2\over\Phi}\right)
+S_m[\bar g_{\mu\nu}, \psi_{(a)}]\,.
$$
In the spirit of \cite{us}, one could therefore introduce 
a ``detuned" version of General Relativity based  on the action, 
$$
S_{\rm detuned GR}
={1\over16\pi}\int d^4x\sqrt{-\bar g}
\left(\Phi\,\bar R+{3\over2}{(\bar\partial\Phi)^2\over\Phi}\right)
+S_m[\Phi F(\Phi)\,\bar g_{\mu\nu}, \psi_{(a)}]\,,
$$ 
which reduces to ``veiled" General Relativity if $F(\Phi)=1$
 and to $\omega=-3/2$ Brans-Dicke theory if $F(\Phi)=\Phi^{-1}$.
 We shall not pursue  this idea any further here.\\
}\\

Let us now be more specific about the matter action $S_m$. 
\smallskip

As an example (others are considered in the Appendix), take matter 
to be an electron characterized by its inertial mass $m$ and 
charge $q$ interacting with the electromagnetic field $A_\mu$ created by 
 an infinitely massive proton,
so that $S_m$ is the Lorentz action where
 $\eta_{\mu\nu}\to g_{\mu\nu}$:\footnote{In Planck units  $m$ and $q$
 are two dimensionless numbers which are determined in a local
 inertial frame where gravity is ``effaced"\cite{Damour87} and
 where the laws of Special Relativity hold.\\}
\begin{equation}
S_m[g_{\mu\nu},L]=-m\int_L\!\sqrt{-g_{\mu\nu}dx^\mu dx^\nu}
+q\int_LA_\mu dx^\mu\,,
 \label{LorentzAction}
\end{equation}
 where $L$ is a path determined by $x^\mu=x^\mu(\lambda)$. 
The equation of motion of the electron, ${\delta S_m\over\delta L}=0$,
  is the familiar Lorentz equation,
\begin{equation}
m\,u^\nu D_\nu u^\mu=q\,F^\mu{}_\nu\,u^\nu\,,
\label{Lorentz}\end{equation}
where $u^\mu={dx^\mu/d\tau}$ with $g_{\mu\nu}u^\mu u^\nu=-1$ and 
$F_{\mu\nu}=\partial_\mu A_\nu-\partial_\nu A_\mu$.

Equivalently, $S_m$ reads, in terms of the metric $\bar g_{\mu\nu}$,
\begin{equation}
S_m[\bar g_{\mu\nu},\Phi, L]
=-\int_L\bar m\sqrt{-\bar g_{\mu\nu}dx^\mu dx^\nu}
+q\int_L\bar A_\mu dx^\mu\,,
\label{veiledLorentzAction}
\end{equation}
where $\bar A_\mu=A_\mu$  and 
\begin{equation}
\bar m=\sqrt{\Phi}\,m\,.
\label{barmass}
\end{equation}
As for the Lorentz equation (\ref{Lorentz}),  it becomes 
\begin{equation}
\bar m\left[\bar u^\nu\bar D_\nu\bar u^\mu
+{1\over2\Phi}\partial_\nu\Phi
\left(\bar g^{\mu\nu}+\bar u^\mu\bar u^\nu\right)\right]
=q\bar F^\mu{}_{\nu}\bar u^\nu\,,
\label{veiledLorentz}
\end{equation}
with $\bar u^\mu={dx^\mu/d\bar\tau}$ 
and $\bar g_{\mu\nu}\bar u^\mu\bar u^\nu=-1$.

In locally Minkowskian coordinates $X^\mu$ in the neighbourhood of
 some point $P$ where $\bar g_{\mu\nu}\approx\eta_{\mu\nu}$ and if 
$\Phi$ is approximately constant, this equation takes the form,
\begin{equation}
\bar m{dU^\mu\over d\tau_M}\approx q\,F^\mu{}_\nu\,U^\nu\,,
\end{equation}
 with $U^\mu={dX^\mu/d\tau_M}$ and $\eta_{\mu\nu}U^\mu U^\nu=-1$. 
This equation is the same as the one governing the motion of the
 electron in Special Relativity apart from the fact that its mass 
is rescaled by the factor $\sqrt{\Phi(P)}$, 
see \cite{Dicke62}.\footnote{This space-time dependence of 
the (inertial) mass  can be  interpreted as a  local rescaling 
 of the unit of mass, see \cite{Dicke62}. It can also be 
interpreted as the result of the ``interaction" of the ``scalar 
field" $\Phi$ with matter.  It must be remembered however that 
this ``interaction" is an artefact of the introduction of
 the metric $\bar{\bm g}$, and that the ``scalar force" 
which appears in (\ref{veiledLorentz}) or (\ref{veiledBianchi})
 can be globally effaced by returning to the original 
metric ${\bm g}$, just like an inertial force can be effaced 
by going to an inertial frame.\\}
 
 As an illustration of the consequences of the rescaling of 
the mass in veiled General Relativity, consider for example 
a transition between the levels $n$ and $n'$ of, say, 
the hydrogen atom.  Its frequency is given by Bohr's formula,
\begin{eqnarray}
\bar \nu(P)
=\left({1\over n'^2}-{1\over n^2}\right){\bar m(P)q^4\over 2}
\qquad\mbox{with}\qquad \bar m(P)=\sqrt{\Phi(P)}\,m\,.
\label{bohr}
\end{eqnarray}
 It  depends on $P$, that is, on when and where it is measured. 
 Hence the frequency $\bar \nu(P)\equiv\bar \nu$ of the
 transition measured at point $P$ (``there and then") and the frequency 
$\bar \nu(P_0)\equiv\bar\nu_0$  of the same transition 
measured at $P_0$ (``here and now") are related
 by:\footnote{This difference between the two numbers
$\bar\nu$ and $\bar\nu_0$ can  be interpreted  as  simply due to
the fact that they are expressed using a different unit of
 time at $P$ and $P_0$, see \cite{Dicke62}.\\} 
 \begin{equation}
\bar\nu=\sqrt{\Phi(P)\over \Phi(P_0)}\,\bar\nu_0\,.
\label{frequency}
\end{equation}

\section{conformal equivalence in cosmology}

Let us show here on a few examples that the standard cosmological 
models of General Relativity or its conformally related sister 
theories all lead to the same physical predictions and hence are 
observationally indistinguishable.\\  

 The field equations to solve are the veiled Einstein
 equations (\ref{veiledEinstein})-(\ref{veiledBianchi})
 for $\bar g_{\mu\nu}$ and $\Phi$. \\
 
We look for simplicity for spatially flat Robertson-Walker 
metrics, 
\begin{eqnarray}
d\bar s^2=\bar a^2(t)(-dt^2+d\vec r^2)\,,
\end{eqnarray}
where the
 scale factor $\bar a$ and the scalar field $\Phi$ depend on 
$t$ only. By construction 
equations (\ref{veiledEinstein})-(\ref{veiledBianchi})
 are undetermined and  we shall choose here, to make our 
point more strikingly, $\Phi$ to be the dynamical field
 describing gravity by imposing
 \begin{eqnarray}
 \bar a(t)=1\,.
 \end{eqnarray}
 Therefore the metric ${\rm\bar g}$ is flat.\footnote{This
 does not mean that $t$ and $\vec r$ represent time and position
 in an inertial frame since the wordlines of free particles
 are not straight lines. They rather solve,
 see  (\ref{veiledLorentz}): 
$\bar u^\nu\bar D_\nu\bar u^\mu
=-{1\over2\Phi}
(\bar\partial^\mu\Phi+\bar u^\mu\bar u^\nu\partial_\nu\Phi)$, 
whose solution is,
 $\vec C$ being three constants: 
$\bar {\vec V}\equiv{\bar{\vec u}/\bar u^0}
={\vec C/\sqrt{C^2+\Phi(t)}}\neq \mbox{const.}$.\\}
 
 Matter is represented by the stress-energy tensor of a perfect 
fluid (see Appendix):
 $\bar T_{\mu\nu}
=(\bar \rho+\bar p)\bar u_{\mu}\bar u_{\nu}+\bar p\,\bar g_{\mu\nu}$
 that we choose to be at rest with respect to the Minkowskian 
coordinate grid $(t, \vec r)$:\footnote{This is the familiar
 ``Weyl postulate".\\} $\bar u^\mu=(1,\vec 0)$~; as for the (veiled)
  density and pressure $\bar\rho$ and $\bar p$ they depend on $t$.
 
The equations of motion (\ref{veiledEinstein})-(\ref{veiledBianchi})
for $\Phi$ then reduce to, a prime denoting derivation with respect to $t$,
\begin{equation}
 {3\over4\Phi}\Phi'^2=8\pi\bar\rho\,,
\qquad \bar\rho'={\Phi'\over2\Phi}(\bar\rho-3\bar p)\,,
\label{veiledFriedmann}
\end{equation}
 which can be solved once an equation of state is given.
 For $\bar p=w\bar \rho$  for example,
\begin{equation}
\Phi=\left(t\over t_0\right)^{4/(1+3w)}\,,
\quad
\bar\rho={3\over2\pi (1+3w)t_0^2}
\left(t\over t_0\right)^{2(1-3w)/(1+3w)}\,.
\end{equation}
 
Let us now turn to the relation between the luminosity distance $D$ 
and redshift $z$ that the model predicts. 

As usual, we focus on a given atomic transition line
in the spectrum of a distant galaxy at point $P=(t,\vec r)$.
 The observer is at point $P_0=(t_0,\vec0)$, and  the
atomic line emitted by this galaxy is observed at frequency $\nu_0$.
 As given in (\ref{frequency}), if $\bar\nu$ is the 
frequency of this transition measured at point $P$,
the frequency of the same transition measured at point 
$P_0$ will be $\bar\nu_0=\sqrt{\Phi(P_0)/\Phi(P)}\,\bar\nu$.
Therefore the observed redshift is given by
\begin{eqnarray}
1+z=\frac{\bar\nu_0}{\nu_0}
=\sqrt{\frac{\Phi(t_0)}{\Phi(t)}}\frac{\bar\nu}{\nu_0}\,.
\label{zform}
\end{eqnarray}

The luminosity distance is given, by definition, as
\begin{eqnarray}
D=\sqrt{\frac{L}{4\pi\ell}}\,,
\label{DZ}
\end{eqnarray}
where $L$ is the absolute luminosity of the galaxy
and $\ell$ is the apparent luminosity per unit area
observed at point $P_0$.
Since the mass of the electron in
 veiled General Relativity varies according to 
$\bar m=\sqrt{\Phi}\,m$, it is crucial here to
 recall that the absolute luminosity
is {\sl not} equal to the luminosity measured 
at the point of emission $P$ (where the frequency
 of the transition is $\bar\nu$) but is defined 
as if the galaxy were at the point of reception
 $P_0$ (where the frequency of the transition is $\bar\nu_0$)
 so that we have
\footnote{This (crucial) coupling of 
the inertial masses  to the scalar field $\Phi$ is forgotten in some 
papers, see e.g. \cite{Capozziello10}, which hence (wrongly) 
conclude to the inequivalence of the Jordan and Einstein frames.\\}
\begin{eqnarray}
L=N\frac{\bar\nu_0}{\Delta t}=N\bar\nu_0^2\,,
\label{abslum}
\end{eqnarray}
where $N$ is the number of photons emitted by this transition
during a period $\Delta t=1/\bar\nu_0$.
The apparent luminosity is given by
\begin{eqnarray}
\ell=N\frac{\nu_0^2}{S}=N\frac{\nu_0^2}{4\pi r^2}\,,
\label{applum}
\end{eqnarray}
where $S=4\pi r^2$ is the surface area of a sphere of
radius $r$ since the metric ${\rm\bar g}$ is flat.
Inserting Eqs.~(\ref{abslum}) and (\ref{applum}) into
(\ref{DZ}), we find, using (\ref{zform}),
\begin{eqnarray}
D={\bar\nu_0\over\nu_0}\,r
=\sqrt{\frac{\Phi(t_0)}{\Phi(t)}}{\bar\nu\over\nu_0}\,\,r\,.  
\label{Dform}
\end{eqnarray}

In order finally to relate $\bar\nu$ to $\nu_0$ and $r$ to $T$
 we must study the propagation of light from $P$ to $P_0$. 
Light follows the null cones of $({\cal M},\bar g_{\mu\nu}=\eta_{\mu\nu})$ 
so that $r$ is the time light takes to propagate from $P$ to $P_0$,
 and the frequency $\bar\nu$ measured at $P$ is the same as 
the frequency $\nu_0$  observed at $P_0$:
\begin{equation}
(\bar\nu=\nu_0\,,\quad r=t_0-t)\qquad
\Longrightarrow\qquad
 z=\sqrt{\Phi(t_0)\over\Phi(t)}-1\,,
\quad D=\sqrt{\Phi(t_0)\over\Phi(t)}\,(t_0-t)\,.\label{DZthree}
\end{equation}
Let us, for cosmetics, trade an integration on $t$ by an integration
 on $z$: 
\begin{eqnarray}
t_0-t=\int_t^{t_0}\! dt=-\int_0^z\!{dz\over dz/dt}
=\int_0^z\!{2\Phi^{3/2}\over\Phi'}dz\,. 
\end{eqnarray}
This leads us to the  relationship between the luminosity-distance 
and redshift that our cosmological model in veiled 
General Relativity predicts:
\begin{equation}
D=(1+z)\int_0^z\!{dz\over H}\,,
\label{DZfinal}
\end{equation}
where $H\equiv\Phi'/(2\Phi^{3/2})$ must be expressed 
in terms of $z=\sqrt{\Phi(t_0)\over\Phi(t)}-1$ after 
integration of the equations of motion (\ref{veiledFriedmann})
 for $\Phi$.\\

Now, in General Relativity, that is, in the ``unveiled frame",
 $ds^2=a^2\,d\bar s^2$ with $a=\sqrt\Phi$, where matter 
is minimally coupled to the metric $g_{\mu\nu}=a^2\eta_{\mu\nu}$,
 $H$ is nothing but the ``Hubble parameter":
\begin{eqnarray}
H\equiv{\Phi'\over2\Phi^{3/2}}
={a'\over a^2}=\frac{1}{a}{d a\over d\tau}\,,
\end{eqnarray}
 with $d\tau\equiv a\, dt$. Moreover the  equations of
 motion (\ref{veiledFriedmann}) for $\Phi$ are identical to the
 standard Friedmann-Lema{\^\i}tre equations,
\begin{eqnarray}
3H^2=8\pi \rho\,,\quad
\dot\rho+3H(\rho+p)=0\,,
\end{eqnarray}
with  $\rho\equiv\bar\rho/\Phi^2$ and $p\equiv\bar p/\Phi^2$
 (see Appendix).
 Finally, the text-book derivation of the relation luminosity-distance 
versus redshift yields (\ref{DZfinal}). Therefore the predicted 
relationship between the observables $z$ and $D$ is the same, 
whether we represent  gravity by a curved Robertson-Walker 
metric $g_{\mu\nu}=a^2\eta_{\mu\nu}$ minimally coupled to matter as
 in General relativity, or by a flat metric $\bar g_{\mu\nu}=\eta_{\mu\nu}$
 together with a scalar field $\Phi$ coupled 
to matter, in its ``veiled" version.\\

The physical {\sl interpretation} of (\ref{DZfinal}) is however 
different. Indeed, in the particular version of veiled General 
Relativity that we considered here:  
\begin{list}{--}{}
\item
The evolution of the universe is not interpreted by cosmic expansion.
 Since we chose $\Phi=a^2$ there is in fact no cosmic expansion
 at all: $\bar g_{\mu\nu}=\eta_{\mu\nu}$~; but we defined on this flat 
manifold a scalar field $\Phi$  which evolves in time and describes
 the interaction of gravity and matter.

\item
There is no redshifting of photons, since the frequency of an 
atomic transition {\sl measured} at $P$ is equal to the frequency
 of that same transition as {\sl observed} at $P_0$ ($\bar\nu=\nu_0$). 

\item
However the interaction of $\Phi$ with matter implies that the 
mass $\bar m$ of the electron varies in time
 ($\bar m=\sqrt\Phi\,m=a\,m$). Therefore the frequency of an
 atomic transition as measured in a lab there and then at $P$ 
is not the same as the frequency measured here and now at
 $P_0$:
 $\bar\nu=\sqrt{\Phi(P)/\Phi(P_0)}\,\bar\nu_0=(a/a_0)\bar\nu_0$.
 This redshifting due to a varying mass is exactly the same
 as the one due to a cosmological redshift in General Relativity.
\end{list}

Pursuing the above interpretation, the temperature of the
cosmic microwave background can be considered constant, since 
photons are not redshifted, and chosen to be the present 
temperature $T_0=2.725$K, throughout the whole history
of the universe (that is, during the whole time-evolution of the
 gravitational field $\Phi$). The universe was in thermal equilibrium when
the electron mass was smaller by a factor of more than $10^3$
compared to the mass today, that is when the ground state 
binding energy of the hydrogen was less than $0.0136\,\mbox{eV}$.
 The ``Big-Bang"  is flat space at time $t=0$ when 
the masses of the matter constituents are zero.\\

In conclusion, the above considerations show that the physical 
interpretation of the equations can be very different in 
General Relativity or its veiled versions, but the resulting 
relations between observables
are completely independent of the conformal representation
 (or ``frame") one chooses.

\section{conformal equivalence in local gravity}

We shall see here that the  tests of General Relativity in 
the Solar System (gravitational redshift, bending of light, 
perihelion advance, Shapiro effect...) can just as well be 
constructed using veiled General Relativity.\\

For definiteness let us describe the gravitational field of 
the Sun by the 
Schwarzschild solution of the vacuum Einstein equations
 written in Droste coordinates $x^\mu=(t,r,\theta,\phi)$,
\begin{equation}
ds^2\equiv g_{\mu\nu}dx^{\mu} dx^{\nu}
=-\left(1-{2M/ r}\right)dt^2+{dr^2\over1-2M/r}
+r^2(d\theta^2+\sin^2\theta d\phi^2)\,,
\label{Schwarzschild}
\end{equation}
where $M$ is the (active) gravitational mass of the Sun. 
The propagation of light and the motion of planets in the
 Solar System are represented by (null) geodesics of this 
Schwarzschild spacetime. Proper time  as measured in, say, 
Planck units, by  a  clock travelling in the Solar System is
 represented by the length of its timelike worldline $x^\mu(\lambda)$,
 that is, by the number $\tau=\int\!\sqrt{-g_{\mu\nu}dx^{\mu} dx^{\nu}}$.\\

Let us now introduce the following ``veiled" Schwarzschild
 line element $ds^2=\Phi\, d\bar s^2$ with $\Phi=1-2M/r$ so that
\begin{equation}
 d\bar s^2\equiv \bar g_{\mu\nu}dx^{\mu} dx^{\nu}
=-dt^2+{dr^2\over(1-2M/r)^2}
+{r^2\over 1-2M/r}(d\theta^2+\sin^2\theta d\phi^2)\,,
\label{veildedSchwarzschild}
\end{equation}
which solves the ``veiled" vacuum Einstein 
equations (\ref{veiledEinstein}) (we shall restrict our
 attention to the region outside the horizon,
 $r>2M$).\footnote{One
 may wonder if the line elements $d\bar s^2=ds^2/\Phi$, $ds^2$ 
being the Schwarzschild solution, are the only solutions of
 the ``veiled" vacuum Einstein equations (which, beware,
 are not $\bar G_{\mu\nu}=0$~!). The answer is yes since,
 by construction, these equations are undetermined and $\Phi$ 
can be chosen at will to solve them. One then chooses $\Phi=1$ 
and invokes the uniqueness of the Schwarzschild solution.\\} 

Light  follows the null geodesics of $\bar g_{\mu\nu}$ which are 
the same as those of $g_{\mu\nu}$. Therefore the prediction for 
the bending of light is the same as in General Relativity.

Test particles do not follow geodesics  of $\bar g_{\mu\nu}$ and
 their equation of motion is given by (\ref{veiledLorentz})
 (with $q=0$). However this equation is just a rewriting of the 
geodesic equation in the metric $g_{\mu\nu}$. Therefore the
 trajectories $r=r(\phi)$ in the equatorial plane $\theta=\pi/2$ 
are the same in both General Relativity and its veiled version.
 The prediction for, say, the perihelion advance of Mercury
 is hence the same.

Consider now an atom at rest at $r$ and an observer at rest at $r_0$.
Since $t$ is proper time, 
the frequency $\nu_0$ of an atomic transition, as observed at $r_0$, 
will be the same as the frequency $\bar\nu$ measured 
at $r$: $\bar\nu=\nu_0$.
 However, in close analogy with the cosmological case,
since the mass of the electron undergoing this 
transition depends on $r$ as Eq.~(\ref{barmass}),
$\bar m=\sqrt{\Phi(r)}\,m$, $\bar\nu$ is related to the frequency
 $\bar\nu_0$ of the transition as measured at $r_0$
by Eq.~(\ref{frequency}): $\bar\nu=\bar\nu_0\sqrt{\Phi(r)/\Phi(r_0)}$.
Hence the gravitational 
redshift is predicted to be
\begin{equation}
z\equiv\frac{\bar\nu_0}{\nu_0}-1
=\sqrt{\Phi(r_0)\over\Phi(r)}-1=\sqrt{1-2M/r_0\over 1-2M/r}-1\,,
\label{gravZ}
\end{equation}
which is exactly the same as the prediction of General Relativity.

Finally let us consider predictions for the tests
of General Relativity 
relying on time measurements (such as the Shapiro effect, GPS,...).
In veiled General Relativity, the proper time interval
$d\bar\tau=\sqrt{-\bar g_{\mu\nu}\,dx^{\mu}dx^{\nu}}$ between
two adjacent events  $x^\mu=(t,r,\theta,\phi)$ and $x^\mu+dx^{\mu}$
differs from that of General Relativity $d\tau$: 
$d\bar\tau=d\tau/\sqrt{\Phi}$. However, if we recall that
time measurements  are based on atomic
clocks, that is, time intervals are counted in units of a frequency 
of an atomic transition, we readily find that the
observed number of `ticks' will be the same,
\begin{eqnarray}
N_{\rm ticks}
=\bar\nu\,d\bar\tau =\sqrt{\Phi}\,\nu\, \frac{d\tau}{\sqrt{\Phi}}
=\nu\, d\tau\,,
\end{eqnarray}
where $\bar\nu$ and $\nu$ are the frequencies of an atomic
transition defined in veiled and unveiled General Relativity,
respectively. Thus predictions for all the time measurements
in veiled General Relativity again exactly agree with
those in General Relativity.

\section{Conclusion}

In 1912 Nordstr\"om proposed a theory where gravity was represented by 
a scalar field $\Phi$ on Minkowski spacetime with metric 
$\bar g_{\mu\nu}=\eta_{\mu\nu}$. Of course, matter was non-minimally 
coupled to that field, so that its interaction to gravity be 
described (see e.g. \cite{Giulini06}).
 In 1914 Einstein and Fokker  introduced a conformally flat 
metric $g_{\mu\nu}=\Phi\,\eta_{\mu\nu}$ which turned Nordstr\"om's equation
 of motion of test particles into the geodesic equation of the metric
 ${\bm g}$. Hence matter was minimally coupled to ${\bm g}$. 
As for the Klein-Gordon field equation for $\Phi$ it became an 
equation relating the scalar curvature of ${\bm g}$ to the trace of 
the stress-energy tensor of matter. It was clear (at least to 
Einstein and Fokker~!) that the two versions of the theory were
 strictly equivalent, Nordstr\"om's formulation being the ``veiled" 
one. And if Nordstr\"om's theory was soon abandoned it was  
not because it had been formulated first in flat spacetime but because
 its predictions (deduced either from its ``veiled" or 
``unveiled" formulations) were in contradiction with observations.
  
  In this paper we did nothing more than what Einstein and Fokker 
did in 1914 but applied the idea to General Relativity itself, in 
order to show, in a hopefully clear way, that, even if the description 
of phenomena could be different in General Relativity and in its
 conformally related sister theories, the predictions for the
 relationships between (classical) observables were strictly the same.
  
   It should then become obvious that the same conclusion holds too 
when dealing with extensions of General Relativity such as $f(R)$
 theories, coupled quintessence or, more generally, scalar-tensor
 theories (even if the scalar field $\Phi$ is then truly dynamical):
 the Jordan frame, where matter is minimally coupled to the metric,
 and the Einstein frame, where the action for gravity is Hilbert's,
 are equivalent, mathematically and physically, at least when dealing
 with classical phenomena and the motion of objects which are weakly
 gravitationally bound. Preferring to interpret the phenomena in the
 Jordan frame is somewhat similar to preferring to work in an inertial
 frame in Special Relativity: this allows to forget about the
 spacetime dependence of the inertial mass of the matter constituents 
just like one can forget about inertial forces in an inertial frame. 
   
This analogy between inertial forces and non-minimal couplings points
 to quantum phenomena where the equivalence between the Jordan and
 Einstein frames may not hold.

Another point which deserves further investigation is the equivalence 
of conformally related frames when it comes to the motion of compact
 bodies whose gravitational binding energy is significant. It is known 
for example that a small black hole follows a geodesic in General
 Relativity \cite{Thorne85}. In scalar tensor theories weakly
 gravitating bodies follow geodesics of the Jordan frame metric 
(to which matter is minimally coupled) but small black holes follow 
geodesics of the Einstein metric, see \cite{Hawking72} and
 e.g. \cite{Shapiro94}. How this result, which is interpreted as a 
violation of the Strong Equivalence Principle, can be obtained
 using the Jordan frame exclusively remains to be elucidated.

\begin{acknowledgments}
N.D. thanks the Yukawa Institute for its hospitality and financial
 support during the YKIS2010 Symposium and Gravity 
and Cosmology 2010 Workshop.
The work of M.S. is supported in part  by 
Korea Institute for Advanced Study under the KIAS Scholar program,  
by the Grant-in-Aid for the Global COE Program 
``The Next Generation of Physics, Spun from Universality and Emergence''
from the Ministry of Education, Culture, 
Sports, Science and Technology (MEXT) of Japan, 
by JSPS Grant-in-Aid for Scientific Research (A) No.~21244033,
and by JSPS Grant-in-Aid for Creative Scientific Research No.~19GS0219. 
\end{acknowledgments}

\appendix
\section{}

We considered in the main text the example of matter being an 
electron in the field of an infinitely massive proton.

 As a second example, consider matter to be a massive scalar field 
$\psi$ with action,
 \begin{equation}
 S^{[\xi]}_m=-{1\over2}\int\! d^4x\sqrt{-g}
\left[(\partial\psi)^2+\left(m^2+{\xi\over6}R\right)\psi^2\right]\,.
 \label{ScalarAction}
 \end{equation}
When $\xi=0$ and $\xi=1$, its extremisation with respect to $\psi$
 yields the familiar Klein-Gordon equations which read, respectively,
\begin{equation}
 \square\psi-m^2\psi=0\,,
\qquad 
\square\psi-\left(m^2+{R\over6}\right)\psi=0\,.
 \label{KleinGordon}
\end{equation}
As for the veiled versions of (\ref{ScalarAction}) the are, 
respectively,
 \begin{equation}
 S^{[0]}_m=-{1\over2}\int\! d^4x\sqrt{-\bar g}\,
\Phi\left[(\bar \partial\psi)^2+\bar m^2\psi^2\right] \,,
\qquad S^{[1]}_m=-{1\over2}\int\! d^4x\sqrt{-\bar g}
\left[(\bar \partial\bar\psi)^2
+\left(\bar m^2+{\bar R\over6}\right)\bar\psi^2\right]
 \label{ScalarVeiledAction}
 \end{equation}
where $g_{\mu\nu}=\Phi\bar g_{\mu\nu}$, $\bar m=\sqrt{\Phi}\,m$, and 
where $\bar\psi=\sqrt{\Phi}\,\psi$ in $S^{[1]}_m$. The 
extremisations of $S^{[0]}_m$ with respect to $\psi$ and of 
$S^{[1]}_m$ with respect to $\bar\psi$ yield, respectively,
\begin{equation}
\bar\square\psi-\bar m^2\psi=-\bar\partial\psi\,\cdot
\,{\bar\partial\Phi\over\Phi}\,,
\qquad  \bar\square\bar\psi
-\left(\bar m^2+{\bar R\over6}\right)\bar\psi=0
\label{veiledKleinGordon}
\end{equation}
which are nothing but a rewriting of equations (\ref{KleinGordon}).
 In locally Minkowskian coordinates $X^\mu$ in the neighbourhood of
 some point $P$ where $\bar g_{\mu\nu}\to\eta_{\mu\nu}$ and where
 $\Phi$ is approximately constant they reduce to their Special 
Relativistic forms, where the mass of the field has to be
 rescaled: $m\to\bar m=m\sqrt{\Phi(P)}$.\footnote{Note 
that the same rescaling of mass occurs 
in a conformal transformation of the Dirac equation,
 see e.g. \cite{BirrellDavies}.\\}

In the case $\xi=0$ the coupling of $\psi$ to $\Phi$ can be 
globally effaced, by returning to the metric $\rm g$.
In the case $\xi=1$ the Klein-Gordon equation is, as is
 well-known, conformally invariant. 

In the conformal invariant case, one might be confused
by the fact that the stability of the field $\psi$
depends on the sign of $(m^2+R/6)$, while one can easily
change its sign by a conformal transformation.
This seemingly paradoxical situation is resolved by
investigating more carefully the relation between the
field in two different conformal frames.

As an example, let us consider the case
 when $g_{\mu\nu}=\eta_{\mu\nu}/(H\eta)^2$ is the (expanding part of)
de Sitter metric (with $-\infty<\eta<0$),
 and $m^2<0$ but $m^2+R/6=m^2+2H^2>0$, so that the field
$\psi$ is stable. Now consider the conformal transformation
to the frame $\bar g_{\mu\nu}=\eta_{\mu\nu}$. Then we have 
$\bar m^2=m^2/(H\eta)^2<0$. Thus the field is badly unstable
because the mass-squared is not only negative but diverges
at $\eta=0$. However if we recall that $\bar\psi=(-H\eta)^{-1}\psi$,
this instability is solely due to the ill behaviour of the 
conformal factor as $\eta\to-0$.

Now let us consider a converse case when $g_{\mu\nu}=\eta_{\mu\nu}$
 and $m^2<0$, so that the field $\psi$ is unstable:
 $\psi\propto e^{|m|\eta}$ diverges exponentially.
Turn now to the expanding de Sitter frame 
$\bar g_{\mu\nu}=\eta_{\mu\nu}/(H\eta)^2$, with $-\infty<\eta<0$.
Then the effective mass-squared $\bar m^2+\bar R/6=m^2/(H\eta)^2+2H^2$
will eventually become positive as $\eta\to-0$, hence the
field must be stable in the expanding de Sitter frame.
This seeming paradox can be resolved by noting the
fact that $\bar\psi=(-H\eta)\psi\propto H\eta e^{|m|\eta}$.
Thus the time is bounded from above at $\eta=0$,
and hence there is literally `no time' for the instability to
develop.\footnote{We thank Andrei Linde for raising this issue.\\}
\\

As a last example consider matter to be a perfect fluid. 
Its stress-energy tensor and equations of motion, deduced from 
their special relativistic expressions by the 
replacement $\eta_{\mu\nu}\to g_{\mu\nu}$ are
\begin{equation}
T_{\mu\nu}=(\rho+p)u_{\mu}u_{\nu}+pg_{\mu\nu}\,,
\qquad D_jT^{\mu\nu}=0\label{Euler}
\end{equation}
where $\rho$ and $p$ are the energy density and pressure of 
the fluid measured in a local inertial frame and where $u^i$
 is its 4-velocity  normalised as $g_{\mu\nu}u^iu^j=-1$. 
Now, since
 $T_{\mu\nu}\equiv-{2\over\sqrt{-g}}{\delta S_m\over\delta g^{\mu\nu}}$ 
(where we need not specify $S_m$), we have 
\begin{eqnarray}
\bar T_{\mu\nu}\equiv-{2\over\sqrt{-\bar g}}
{\delta S_m\over\delta \bar g^{\mu\nu}}=\Phi T_{\mu\nu}\,, 
\end{eqnarray}
so that the ``veiled" version of (\ref{Euler}) is,
 cf (\ref{veiledBianchi}),
\begin{equation}
\bar T_{\mu\nu}
=(\bar \rho+\bar p)\bar u_{\mu}\bar u_{\nu}+\bar p\,\bar g_{\mu\nu}\,,
\qquad \bar D_{\nu}\bar T^{\mu\nu}={\bar\partial^{\mu}\Phi\over2\Phi}\,
\bar T\,,\label{veiledEuler}
\end{equation}
where $\bar g_{\mu\nu}\bar u^i\bar u^j=-1$, and with
 $\bar\rho=\Phi^2\rho$ and $\bar p=\Phi^2 p$.\footnote{Since 
the dimensions of $\rho$ and $p$ are $[M][L]^{-3}$ their 
rescaling is in keeping with the local rescaling of units 
alluded to in a previous footnote:
 $[M]\to[\bar M]=\sqrt{\Phi}[M]$ and $[L]\to[\bar L]=[L]/\sqrt{\Phi}$.\\}

\providecommand{\href}[2]{#2}\begingroup\raggedright
\endgroup


\begin{thebibliography}{1}

\bibitem{Tsujikawa10} 
  A.~De Felice and S.~Tsujikawa,
  Living Rev.\ Rel.\  {\bf 13}, 3 (2010)
  [arXiv:1002.4928 [gr-qc]].

\bibitem{Copeland06}
  E.~J.~Copeland, M.~Sami and S.~Tsujikawa,
  Int.\ J.\ Mod.\ Phys.\  D {\bf 15}, 1753 (2006)
  [arXiv:hep-th/0603057].

\bibitem{Maeda02}
  Y.~Fujii and K.~Maeda,
 ``The scalar-tensor theory of gravitation,''
 {\it  Cambridge, USA: Univ. Pr. (2003) 240 p}

\bibitem{Damour92} 
  T.~Damour and G.~Esposito-Farese,
  Class.\ Quant.\ Grav.\  {\bf 9}, 2093 (1992).

\bibitem{Gunzig98}
  V.~Faraoni, E.~Gunzig and P.~Nardone,
  Fund.\ Cosmic Phys.\  {\bf 20}, 121 (1999)
  [arXiv:gr-qc/9811047].

\bibitem{Dicke62} 
 R.~H. Dicke,
 Phys. Rev. 125 (1962) 2163.

\bibitem{Flanagan04}
  E.~E.~Flanagan,
  Class.\ Quant.\ Grav.\  {\bf 21}, 3817 (2004)
  [arXiv:gr-qc/0403063].

\bibitem{Catena06} 
  R.~Catena, M.~Pietroni and L.~Scarabello,
  Phys.\ Rev.\  D {\bf 76}, 084039 (2007)
  [arXiv:astro-ph/0604492].
\\
  R.~Catena, M.~Pietroni and L.~Scarabello,
  J.\ Phys.\ A  {\bf 40}, 6883 (2007)
  [arXiv:hep-th/0610292].

\bibitem{Faraoni06}
  V.~Faraoni and S.~Nadeau,
  Phys.\ Rev.\  D {\bf 75}, 023501 (2007)
  [arXiv:gr-qc/0612075].

\bibitem{BransDicke61}
 C. Brans and R.H. Dicke,
 Phys. Rev. 124 (1961) 925.

\bibitem{Dabrowski05}
  M.~P.~Dabrowski, T.~Denkiewicz and D.~Blaschke,
  Annalen Phys.\  {\bf 16}, 237 (2007)
  [arXiv:hep-th/0507068].


\bibitem{us}
  N.~Deruelle, M.~Sasaki and Y.~Sendouda,
  Phys.\ Rev.\  D {\bf 77}, 124024 (2008)
  [arXiv:0803.2742 [gr-qc]].

\bibitem{Damour87}
 T. Damour, 
 in ``Three Hundred Years of Gravitation",
  eds. S.~W. Hawking and W. Israel, 128-198
 (Cambridge University Press, 1987).

\bibitem{Capozziello10} 
  S.~Capozziello, P.~Martin-Moruno and C.~Rubano,
  Phys.\ Lett.\  B {\bf 689}, 117 (2010)
  [arXiv:1003.5394 [gr-qc]].

\bibitem{BirrellDavies} 
 N.~D. Birell and P.~C.~W. Davies,
 ``Quantum field theory in curved space", 
 (Cambridge University Press, 1982).

\bibitem{Giulini06}
  D.~Giulini,
  arXiv:gr-qc/0611100.

\bibitem{Thorne85}
  K.~S.~Thorne and J.~B.~Hartle,
  Phys.\ Rev.\  D {\bf 31}, 1815 (1984).


\bibitem{Hawking72} 
  S.~W.~Hawking,
  Commun.\ Math.\ Phys.\  {\bf 25}, 167 (1972).

\bibitem{Shapiro94} 
  M.~A.~Scheel, S.~L.~Shapiro and S.~A.~Teukolsky,
  Phys.\ Rev.\  D {\bf 51}, 4236 (1995)
  [arXiv:gr-qc/9411026].

\end{thebibliography}
\end{document}